\documentclass[aps,prd,nopacs,floatfix,notitlepage,nofootinbib,twocolumn,a4paper,longbibliography]{revtex4-1}

\usepackage{amsfonts,amsmath,units,wasysym,epsfig,graphicx,verbatim,color,subfigure,graphicx,bm,mathrsfs,lipsum,hyperref,cleveref}
\usepackage{booktabs}
\usepackage[normalem]{ulem}  
\usepackage{multirow}
\usepackage[utf8]{inputenc}
\usepackage{graphicx}
\usepackage{subcaption} 

\begin{document}

\newcommand{\bhaskar}[1]{\textcolor{blue}{ \bf BB: #1}}

\newcommand{\HU}{{Hamburger Sternwarte, Gojenbergsweg 112, D-21029 Hamburg, Germany}}

\title{Equation of State Extrapolation Systematics: Parametric vs. Nonparametric Inference of Neutron Star Structure}

\author{Bhaskar Biswas}
\affiliation{\HU}

\begin{abstract}
The equation of state (EOS) of cold dense matter remains one of the central open problems in nuclear astrophysics. Its inference is complicated by the lack of \textit{ab initio} theoretical control above about twice nuclear saturation density, where the EOS must be extrapolated. Parametric schemes such as piecewise polytropes (PP) are computationally efficient but impose restrictive functional forms, while nonparametric approaches such as Gaussian processes (GP) offer greater flexibility at the cost of larger prior volumes. In this work, we extend our hybrid EOS framework by replacing the high-density polytropic extension with a Gaussian Process representation of the squared sound speed, anchored at low densities by the SLy crust EOS and a nuclear meta-model constrained by $\chi$EFT and laboratory measurements. Using a hierarchical Bayesian analysis, we jointly constrain the EOS and the neutron star mass distribution with multi-messenger data, including NICER radius measurements, GW170817 and GW190425 tidal deformabilities, $2M_\odot$ pulsars, and neutron skin experiments. We explore four scenarios defined by the choice of high-density extrapolation (PP vs.\ GP) and hotspot geometry in the NICER modeling of PSR~J0030$+$0451 (ST+PDT vs.\ PDT-U). We find that GP extrapolations generally yield softer EOS posteriors with broader uncertainty bands. Hotspot geometry assumptions also play an important role, leading to systematic shifts in the inferred mass--radius relations. Bayesian evidence strongly favors the ST+PDT geometry over PDT-U across both extrapolation schemes, while the GP extension is consistently preferred over PP, with substantial support. Taken together, these results underscore the importance of both observational modeling choices and EOS extrapolation strategies in shaping neutron star EOS inferences, and demonstrate that a GP-based extension provides a robust framework for quantifying systematic uncertainties in high-density matter.
\end{abstract}

\maketitle

\section{Introduction}

Over the past decade, \textit{multi-messenger} observations have revolutionized our understanding of the neutron star (NS) equation of state (EOS). Gravitational-wave (GW) detections of binary neutron star (BNS) mergers (GW170817~\cite{TheLIGOScientific:2017qsa,Abbott:2018wiz,Abbott:2018exr}, GW190425~\cite{LIGOScientific:2020aai}) have constrained tidal deformabilities at intermediate densities, while NICER X-ray timing~\cite{riley19c,Riley:2021pdl,Miller:2019cac,Miller:2021qha,Choudhury:2024xbk,Mauviard:2025dmd} has provided complementary mass--radius measurements. Observations of $\sim 2M_\odot$ pulsars~\cite{Antoniadis:2013pzd,Cromartie:2019kug} have ruled out soft EOSs unable to support such high maximum masses. Laboratory measurements of neutron-skin thicknesses (PREX-II~\cite{Reed:2021nqk}, CREX~\cite{CREX:2022kgg}) have further refined constraints on the symmetry energy near nuclear saturation density. On the theory side, chiral effective field theory ($\chi$EFT)~\cite[see, e.g.,][]{Epelbaum:2008ga,Machleidt:2011zz,Hammer:2012id,Hebeler:2020ocj,Drischler:2021kxf} provides a controlled description of nuclear interactions up to densities of $\sim 1.5$--$2 n_s$, where $n_s \approx 0.16~\mathrm{fm}^{-3}$ denotes the nuclear saturation density, while perturbative QCD~\cite{Gorda:2021znl,Gorda:2022jvk,Komoltsev:2021jzg} is reliable only at extremely high densities ($\sim 40 n_s$). This leaves a wide intermediate-density regime that must be bridged phenomenologically.

A central challenge in EOS inference therefore lies in \textbf{extrapolating the EOS beyond the domain of theoretical control}. Parametric representations---such as piecewise polytropes (PP)~\cite{Read:2008iy}, 
spectral expansions~\cite{Lindblom:2010bb}, and speed-of-sound 
parameterizations~\cite{Greif:2018njt}---are widely used because they are 
computationally efficient, but their restrictive functional forms can bias the 
inferred pressure--density relation~\cite{Legred:2022pyp} and associated neutron star observables. Complementary approaches based on nuclear metamodels~\cite{Montefusco:2024xrx} 
or relativistic mean-field (RMF) interactions~\cite{Traversi:2020aaa,Passarella:2025zqb} 
retain an explicit connection to nuclear microphysics, but likewise require 
assumptions~\cite{Biswas:2020puz} when extrapolated to supranuclear densities relevant for neutron 
star interiors. By contrast, \textbf{nonparametric approaches}, such as Gaussian Process (GP) representations~\cite{Landry:2018prl,Landry_2020PhRvD.101l3007L,Gorda:2022jvk,Sarin:2023tgv}, treat the EOS as a distribution over functions rather than a fixed template. These methods allow greater flexibility and broader prior support, but they can require larger datasets for meaningful inference and may obscure direct physical interpretation.

To balance these competing strengths and limitations, \textbf{hybrid EOS frameworks} have been developed~\cite{Biswas:2020puz,Biswas:2020xna,Biswas:2021yge}, in which the EOS at low densities is anchored by a nuclear meta-model~\cite{Bombaci:1991zz} informed by empirical constraints (e.g., PREX/CREX, $\chi$EFT, and symmetry energy parameters), while the high-density regime is extended using a phenomenological extrapolation. This construction enables systematic comparison of modelling assumptions while retaining a firm physics-based foundation near nuclear saturation density.

In this work, we extend our previous EOS inference framework~\cite{Biswas:2024hja} in two important ways.  
First, we replace the piecewise polytropic high-density extrapolation with a Gaussian Process representation of the squared sound speed. This allows us to quantify the systematic errors introduced by restrictive parametric forms and assess their impact on the recovered pressure--density relation and neutron star observables. Second, we modify the crust--core junction procedure: 
whereas our earlier work imposed continuity of the 
pressure at a single matching density between the 
crust and the empirical meta-model, here we adopt a 
thermodynamically consistent finite-interval blending 
construction in which only the pressure is interpolated, 
while the chemical potential and energy density are 
reconstructed from the fundamental relations for cold 
barotropic matter. The center of the 
blending region is set to an empirically estimated 
crust--core transition density rather than an ad hoc 
fixed value (see Sec.~\ref{subsection:Crust--core blending}).
This procedure allows for a smoother 
transition between crust and core models while preserving 
the thermodynamic relations across the matching region. By integrating constraints from multi-messenger astrophysics (NICER, gravitational waves, pulsar masses), nuclear theory, and laboratory experiments within a hierarchical Bayesian framework, we provide a systematic comparison of parametric and nonparametric EOS inference and establish robust, bias-minimized constraints on the dense matter EOS. 

\section{EOS Model with Gaussian Process Extension}
\label{section: EOS-model}

Our EOS construction combines three components: (i) a fixed SLy~\cite{Douchin:2001sv} crust at low densities, (ii) a nuclear meta-model (see Sec.~\ref{subsection:meta-model}) anchored near saturation density, characterized by the symmetry energy slope $L$ and curvature $K_{\text{sym}}$, and (iii) a high-density extension governed by a Gaussian Process representation of the squared sound speed.

These components are joined in sequence as follows:  
\begin{itemize}
    \item for $n < n_{cc}$, the EOS is given by the fixed SLy crust;  
    \item for $n_{cc} \leq n \leq n_t$, the EOS is described by the empirical meta-model;  
    \item for $n > n_t$, the EOS is extended using the GP representation.  
\end{itemize}
Here $n_{cc}$ denotes the crust--core transition density (treated with a smooth blending construction, see Sec.~\ref{subsection:Crust--core blending}), while $n_t$ is the matching density at which the empirical meta-model is joined to the GP extension.

\subsection{Nuclear Meta-Model}
\label{subsection:meta-model}

The empirical meta-model \citep{PhysRevC.44.1892} (see also
\cite{Montefusco:2026jlq} for a general discussion of the
meta-model framework) expresses the energy per
nucleon of nuclear matter as a function of baryon density $n$ and
isospin asymmetry $\delta = 1 - 2Y_e$,
\begin{equation}
e(n, \delta) = e_0(n) + e_{\mathrm{sym}}(n)\,\delta^2 + \mathcal{O}(\delta^4),
\label{eq:eos_meta}
\end{equation}
where the isoscalar part $e_0(n)$ and the symmetry energy
$e_{\mathrm{sym}}(n)$ are expanded around the saturation density
$n_0$ as
\begin{eqnarray}
e_0(n) &=& e_0(n_0) + \frac{K_0}{2}\chi^2 + \ldots, \label{eq:e0} \\
e_{\mathrm{sym}}(n) &=& e_{\mathrm{sym}}(n_0) + L\chi +
\frac{K_{\mathrm{sym}}}{2}\chi^2 + \ldots, \label{eq:esym}
\end{eqnarray}
with $\chi = (n - n_0)/(3n_0)$. The coefficients are the standard nuclear matter parameters (NMPs):
$e_0(n_0)$ is the binding energy at saturation, $K_0$ is the
incompressibility, $e_{\mathrm{sym}}(n_0)$ is the symmetry energy,
and $L$ and $K_{\mathrm{sym}}$ are its slope and curvature. The
$\beta$-equilibrated EOS is obtained by imposing charge neutrality
and the condition of chemical equilibrium under weak interactions,
$\mu_n - \mu_p = \mu_e$, which determines the proton fraction $Y_e$
as a function of density. The pressure and energy density are then
given by $P = n^2\,\partial e/\partial n$ and $\varepsilon = n(m_N +
e)$, respectively, where $m_N = 931.494~\mathrm{MeV}$ is the
nucleon mass, evaluated along the $\beta$-equilibrium track,
yielding the EOS as a function of baryon density $n$ alone.

The lower-order NMPs are well constrained by nuclear experiments
and are fixed following our previous work~\cite{Biswas:2021yge}.
Specifically, we fix the saturation density $n_0 = 0.16~\mathrm{fm}^{-3}$
and the energy per nucleon at saturation $e_0(n_0) = -15.9~\mathrm{MeV}$.
A survey of 53 experimental results~\cite{Oertel:2016bki} found
$e_{\mathrm{sym}}(n_0) = 31.7 \pm 3.2~\mathrm{MeV}$ and
$K_0 = 240 \pm 30~\mathrm{MeV}$. Previous Bayesian analyses combining
gravitational-wave and X-ray observations found that current
multi-messenger data do not provide significant constraints on these
parameters~\cite{Biswas:2020puz,Biswas:2021yge}, so we fix $e_{\mathrm{sym}}(n_0)$ and $K_0$ at their survey median values of
$31.7~\mathrm{MeV}$ and $240~\mathrm{MeV}$, respectively, rather than
unnecessarily increasing the dimensionality of the EOS parameterization.
Since the present data do not favor or constrain a particular range of
$e_{\mathrm{sym}}(n_0)$ and $K_0$, our results remain robust under
reasonable variations of these parameters. This leaves the slope $L$
and the curvature $K_{\mathrm{sym}}$ of the symmetry energy as the two
free parameters of the meta-model, both sampled with broad uniform
priors as listed in Table~\ref{tab:prior}.

\subsection{Crust--core blending}
\label{subsection:Crust--core blending}

In our earlier work~\cite{Biswas:2024hja,Biswas:2025ivu}, 
the crust--core junction was determined by enforcing 
continuity of the pressure at a single matching density. 
While this prescription yields a thermodynamically 
consistent EOS, it generally leads to a discontinuity in 
the derivative $dP/d\epsilon$, and hence in the squared 
speed of sound $c_s^2$, at the matching point. A finite-
interval connection between crust and core models can 
provide a smoother transition, but requires care: 
interpolating multiple thermodynamic quantities with a 
density-dependent blending function can lead to 
violations of first-law consistency, including the 
identity $c_s^2 = dP/d\epsilon$. In the present analysis, 
we therefore adopt a thermodynamically consistent 
finite-interval matching procedure in which only the 
pressure is interpolated across the blending region, 
while the chemical potential and energy density are 
reconstructed from the fundamental relations for cold 
barotropic matter~\cite{Fortin:2016hny}.

Instead of performing a single-point match, we connect the SLy crust
and the empirical meta-model across a finite density interval centered
on the crust--core transition density. Rather than fixing $n_{cc}$ to
an ad hoc value, we estimate it self-consistently from the symmetry
energy parameters using the empirical fit of \citet{Ducoin:2011fy},
\begin{equation}
n_{cc} = 3.23 \times 10^{-4}\,\bigl(L_{01} + 0.426\,K_{\mathrm{sym},01}\bigr)
         + 0.0802 \quad [\mathrm{fm}^{-3}],
\label{eq:ncc_ducoin}
\end{equation}
where $L_{01}$ and $K_{\mathrm{sym},01}$ are the symmetry energy slope
and curvature evaluated at $n_{\mathrm{ref}} = 0.1~\mathrm{fm}^{-3}$.
This relation was derived from the crossing of the dynamical spinodal
with the $\beta$-equilibrium line across a wide range of Skyrme and
relativistic nuclear models \citep{Ducoin:2011fy}, and yields transition
densities in the physically expected range
$0.06$--$0.10~\mathrm{fm}^{-3}$, consistent with more microscopic
extended Thomas--Fermi calculations \citep{Klausner:2025ucq}.
In our meta-model, where
$e_{\mathrm{sym}}(n) = e_{\mathrm{sym},0} + L\chi
+ \tfrac{1}{2}K_{\mathrm{sym}}\chi^{2}$
with $\chi = (n - n_{0})/(3n_{0})$, these coefficients reduce to
\begin{equation}
L_{01} = L + K_{\mathrm{sym}}\,\chi_{\mathrm{ref}},
\qquad
K_{\mathrm{sym},01} = K_{\mathrm{sym}},
\label{eq:L01_Ksym01}
\end{equation}
where $\chi_{\mathrm{ref}} = (n_{\mathrm{ref}} - n_{0})/(3n_{0})$.
Since $L$ and $K_{\mathrm{sym}}$ are free parameters in our inference,
$n_{cc}$ varies self-consistently across the posterior rather than
being fixed at a single value.
The blending half-width is held fixed at
$\Delta n = 0.004~\mathrm{fm}^{-3}$, which is a purely numerical
smoothing parameter chosen small enough to have no impact on the
inferred EOS.
We define the blending region
$n_{b} \in [n_{cc} - \Delta n,\, n_{cc} + \Delta n]$.

Within this interval, we construct a smooth ($C^\infty$) weight function
following the prescription of Ref.~\cite{Burrello:2025jay},
\begin{equation}
w(n_b) =
\frac{f(x)}{f(x)+f(1-x)},
\qquad
x = \frac{n_b-(n_{cc}-\Delta n)}{2\Delta n},
\end{equation}
where
\begin{equation}
f(x) =
\begin{cases}
\exp(-1/x), & x>0, \\
0, & x \le 0.
\end{cases}
\end{equation}
This function satisfies $w=0$ at $n_{cc}-\Delta n$,
$w=1$ at $n_{cc}+\Delta n$, and is infinitely differentiable
inside the interval.

We blend only the pressure,
\begin{equation}
P_{\rm blend}(n_b)
=
(1-w)\,P_{\rm SLy}(n_b)
+
w\,P_{\rm emp}(n_b),
\end{equation}
and reconstruct the remaining thermodynamic quantities
from the fundamental relations for cold barotropic matter,
as advocated in Ref.~\cite{Fortin:2016hny}.
In particular, the chemical potential is obtained from
\begin{equation}
\mu(n_b)
=
\mu(n_1)
+
\int_{n_1}^{n_b}
\frac{1}{n}\,
\frac{dP_{\rm blend}}{dn}
\, dn,
\end{equation}
where $n_1 = n_{cc}-\Delta n$.
The energy density then follows from
\begin{equation}
\epsilon(n_b)
=
n_b\,\mu(n_b)
-
P_{\rm blend}(n_b).
\end{equation}

To ensure continuity of the chemical potential at
$n_{cc}+\Delta n$, we apply a constant shift
$\Delta\mu$ to the core-side energy density,
following the matching prescription of
Ref.~\cite{Fortin:2016hny}.
This guarantees continuity of $P$, $\mu$, and
$\epsilon$ across the matching region.

The squared speed of sound is subsequently computed
consistently from
\begin{equation}
c_s^2 = \frac{dP}{d\epsilon}.
\end{equation}

This construction preserves thermodynamic consistency by
building the EOS from a single interpolated pressure function,
while the $C^\infty$ weight ensures a smooth transition
between crust and core models.
Compared to a single-point matching procedure,
it significantly reduces numerical artifacts in the stiffness
of the EOS and improves stability when sampling over wide
ranges of $(L, K_{\rm sym})$.

\subsection{Gaussian Process high-density extension}

Beyond the matching density $n_t$, the squared sound speed $c_s^2(n)$ is modeled as a Gaussian Process (GP), following Refs.~\cite{Landry:2018prl,Essick:2019ldf,Gorda:2022jvk}. To map the causal domain $c_s^2 \in [0,1]$ to the real line, we introduce an auxiliary variable
\begin{equation}
    \phi(n) = -\log\!\left(\frac{1}{c_s^2(n)} - 1\right).
\end{equation}
The GP prior is then placed on $\phi(n)$,
\begin{equation}
    \phi(n) \sim \mathcal{N}\!\left(-\ln\!\left(\frac{1}{\bar c_s^2} - 1\right),\, K(n,n')\right),
\end{equation}
with a squared-exponential kernel
\begin{equation}
    K(n,n') = \eta \exp\!\left[-\frac{(n-n')^2}{2 \ell^2}\right],
\end{equation}
where $\eta$ controls the variance of fluctuations and $\ell$ sets the correlation length in density. The mean function corresponds to the logit transform of a reference sound speed $\bar c_s^2$.  

The GP hyperparameters are drawn hierarchically from broad normal priors,
\begin{equation}
\begin{aligned}
    \ell &\sim \mathcal{N}(1.0 n_s, (0.25 n_s)^2), \\
    \eta &\sim \mathcal{N}(1.25, 0.2^2), \\
    \bar c_s^2 &\sim \mathcal{N}(0.5, 0.25^2),
\end{aligned}
\end{equation}
ensuring that the ensemble of EOSs spans a wide range of plausible supranuclear behaviors. At the matching density $n_t$, the GP is anchored to the meta-model prediction, ensuring continuity of the EOS.

From each realization of $\phi(n)$ we reconstruct the EOS by inverting to obtain $c_s^2(n)$, integrating the baryon chemical potential, and computing the energy density and pressure:
\begin{align}
    \mu(n) &= \mu_t \exp\!\left(\int_{n_t}^n \frac{c_s^2(n')}{n'}\, dn'\right), \\
    \epsilon(n) &= \epsilon_t + \int_{n_t}^n \mu(n') \, dn', \\
    P(n) &= n\mu(n) - \epsilon(n).
\end{align}

This GP construction automatically enforces causality and thermodynamic stability, while providing much greater flexibility than polytrope-based models. By comparing inferences under the two frameworks, we can directly quantify the systematic errors introduced by parametric versus nonparametric extrapolations of the EOS at high densities.

\section{Inference Methodology}
\label{sec:inference}

Following our previous work~\cite{Biswas:2024hja}, we employ a hierarchical Bayesian framework to jointly constrain the neutron star EOS and the mass distribution. This approach integrates diverse datasets spanning theory, experiment, and observation. The posterior distribution of the model parameters is obtained via Bayes’ theorem:
\begin{equation}
P(\boldsymbol{\theta} \mid D) = \frac{P(D \mid \boldsymbol{\theta})\, P(\boldsymbol{\theta})}{P(D)},
\end{equation}
where $\boldsymbol{\theta}$ represents the set of model parameters—including EOS parameters ($\theta_{\rm EOS}$) and mass distribution parameters ($\theta_{\rm pop}$)—and $D$ denotes the combined dataset. The likelihood $P(D \mid \boldsymbol{\theta})$ encodes the agreement between model predictions and data, while $P(\boldsymbol{\theta})$ represents prior knowledge or theoretical constraints. The evidence $P(D)$ serves as a normalization constant and is not required for parameter estimation. A complete summary of the model parameters and their prior ranges is provided in Table~\ref{tab:prior}.

\begin{table}[ht!]
\begin{center}
\caption{Prior ranges of the NS EOS and mass distribution model hyperparameters used in this work. The notation $U$ and $\mathcal{N}$ stands for uniform and normal distribution, respectively.}
\label{tab:prior}
\begin{tabular}{ |c|c|c|c|c| } 
  \hline
  Model & Parameters & Units & Prior \\  
  \hline\hline
  \multirow{7}{3.2em}{EOS}  
                         & $L$ & MeV &   $U(0,150)$\\
                         & $K_{\rm sym}$ & MeV &  $U(-600,100)$\\
                         & $n_{t}$ & $n_s$  & $U(1,2)$\\ 
                         & $l$ & $n_s$   & $\mathcal{N}(1 ,0.25)$\\
                         & $\eta$ & - & $\mathcal{N}(1.25 ,0.2)$\\
                         & $\bar c_s^2$ &  -  & $\mathcal{N}(0.5,0.25)$\\
                         
  \hline
  \multirow{4}{3em}{Mass} & $\mu_1$ & $M_{\odot}$ &  $U(0.9,\mu_2)$\\ 
                          & $\sigma_1$ & $M_{\odot}$  &   $U(0.01,\sigma_2)$\\
                          & $\mu_2$ & $M_{\odot}$ &  $U(0.9,M_{\text{max}})$\\ 
                          & $\sigma_2$ & $M_{\odot}$  &   $U(0.01,1.0)$\\
                          & $w$ & -  &   $U(0.1,0.9)$\\
  \hline
\end{tabular}
\end{center}
\end{table}

\subsection{EOS Model Parameters}
The EOS hyperparameters considered in this work are:
\begin{itemize}
    \item $L$: slope of the symmetry energy at saturation density,
    \item $K_{\text{sym}}$: curvature of the symmetry energy,
    \item $n_t$: transition density between the empirical meta-model and the GP EOS,
    \item $\ell, \eta, \bar c_s^2$: hyperparameters of the GP EOS, corresponding to the correlation length, variance, and reference sound speed squared.
\end{itemize}

In our Bayesian framework, we additionally treat the random seed used to generate Gaussian Process realizations as a parameter. Specifically, we assign a uniform prior over integer values between 1 and 10,000. This choice ensures that the stochasticity inherent to GP draws is explicitly tracked within the inference process. By retaining the seed as part of the posterior, any EOS sample can be uniquely reconstructed, which in turn allows us to evaluate other macroscopic neutron star properties in a consistent and reproducible manner.

\subsection{Mass Distribution Parameters} 
We describe the neutron star mass distribution using a two-component Gaussian mixture, capturing the presence of distinct sub-populations. The model is characterized by:  
\begin{itemize}
    \item \( \mu_1, \mu_2 \): mean masses of the two Gaussian components,  
    \item \( \sigma_1, \sigma_2 \): corresponding standard deviations,  
    \item \( w \): mixture weight of the first component, with the second assigned a weight of \(1-w\).  
\end{itemize}

In compact form, the probability density for a neutron star mass \(M\) is written as
\begin{multline}
        P(M) = \left[w \mathcal{N}(M|\mu_1,\sigma_1)/B 
        + (1-w) \mathcal{N}(M|\mu_2,\sigma_2)/C\right] \\
        \times U(M|M_{\rm min},M_{\rm max}) , \label{bimodal}
\end{multline}
where \(\mathcal{N}(M \mid \mu, \sigma^2)\) denotes a normal distribution with mean \(\mu\) and variance \(\sigma^2\) and the term $U(M|M_{\rm min},M_{\rm max})$ reads:
\begin{equation}
    U(M|M_{\rm min},M_{\rm max}) = 
    \begin{cases} 
        \frac{1}{M_\mathrm{max} - M_\mathrm{min}} & \text{if } M_\mathrm{min} \leq M \leq M_\mathrm{max}, \\
        0 & \text{else.}
    \end{cases}
\end{equation}

\subsection{Likelihood Components}
The likelihood function is constructed by combining the following datasets and theoretical inputs:
\begin{itemize}
    \item \textbf{NICER:} Simultaneous mass–radius inferences for PSRs J0030+0451~\cite{Vinciguerra:2023qxq,vinciguerra_2023_8239000}, J0740+6620~\cite{Salmi:2024aum,salmi_2024_10519473}, J0437--4715~\cite{Choudhury:2024xbk,choudhury_2024_13766753}, and and PSR J0614$-$3329~\cite{Mauviard:2025dmd}.
    \item \textbf{Gravitational waves:} Tidal deformability posteriors from GW170817~\cite{Abbott:2018wiz} and GW190425~\cite{LIGOScientific:2020aai}.
    \item \textbf{Theory priors:} Chiral effective field theory (\(\chi\)EFT) at low densities~\cite{BUQEYEgithub,Drischler:2020hwi}, and perturbative QCD (pQCD) at asymptotically high densities~\cite{Gorda:2021znl,Gorda:2022jvk,Komoltsev:2021jzg}.
    \item \textbf{Laboratory experiments:} Neutron skin thickness measurements from PREX-II~\cite{Adhikari:2021phr} and CREX~\cite{CREX:2022kgg}, which are directly related to the symmetry energy.
    \item \textbf{Radio pulsar masses:} Three classes of neutron star mass measurements~\cite{Alsing:2017bbc,Fan:2023spm}. These include precise individual masses (from Shapiro delay or relativistic orbital parameters), as well as binary systems with total mass or mass ratio constraints combined with the mass function, marginalizing over inclination.
\end{itemize}

We do not include the mass–radius data for PSR J1231$-$1411 to provide a cleaner comparison with earlier works. As shown by~\cite{Salmi:2024bss}, the Bayesian analysis of this system is sensitive to radius priors and exhibits convergence issues, motivating its exclusion here.

\subsection{Full Likelihood}
Combining all ingredients, the full likelihood used in this work can be expressed schematically as
\begin{align}
    \mathcal{L}(D \mid \theta_{\rm EOS}, \theta_{\rm pop}) = \, &
    \mathcal{L}_{\rm GW}(D_{\rm GW} \mid \theta_{\rm EOS}, \theta_{\rm pop}) \,
    \nonumber \\[4pt]
    & \times \mathcal{L}_{\rm NICER}(D_{\rm NICER} \mid \theta_{\rm EOS}, \theta_{\rm pop}) \,
    \nonumber \\[4pt]
    & \times \mathcal{L}_{\rm Mass}(D_{\rm Radio} \mid \theta_{\rm EOS}) \,
    \nonumber \\[4pt]
    & \times \mathcal{L}_{\chi{\rm EFT}}(\theta_{\rm EOS}) \,
    \mathcal{L}_{\rm pQCD}(\theta_{\rm EOS}) \,
    \nonumber \\[4pt]
    & \times \mathcal{L}_{\rm Skin}(D_{\rm skin} \mid \theta_{\rm EOS}) \, .
\end{align}

We note that both $\chi$EFT calculations and neutron-skin 
measurements primarily constrain the EOS of isospin-symmetric 
or pure neutron matter, rather than matter in 
$\beta$-equilibrium. In the present analysis, these inputs are 
therefore incorporated as approximate likelihoods on the 
underlying empirical nuclear parameters, which in turn 
determine the $\beta$-equilibrated EOS. As such, their impact 
on neutron star observables is mediated through the assumed 
isospin dependence of the EOS near saturation density. Detailed derivations and discussions of each likelihood component can be found in~\cite{Biswas:2024hja}. Posterior samples are drawn using a nested sampling algorithm implemented in \texttt{PyMultiNest}~\cite{Buchner:2014nha}.

\section{Results}
\label{section: results}

To quantify the impact of EOS extrapolation strategy and NICER hotspot geometry assumptions, we analyze four scenarios: (i) GP high-density extension with J0030 modeled under the ST+PDT hotspot geometry, (ii) GP extension with the PDT-U geometry, (iii) PP high-density extension with ST+PDT, and (iv) PP extension with PDT-U. In what follows, we use ``GP'' and ``PP'' to specifically refer to the choice of high-density extrapolation, while ``ST+PDT'' and ``PDT-U'' indicate the assumed hotspot geometry of J0030’s X-ray pulse profile.  

\subsection{Impact of modeling assumptions on EOS parameters}

\begin{figure}
    \centering
    \includegraphics[width=0.45\textwidth]{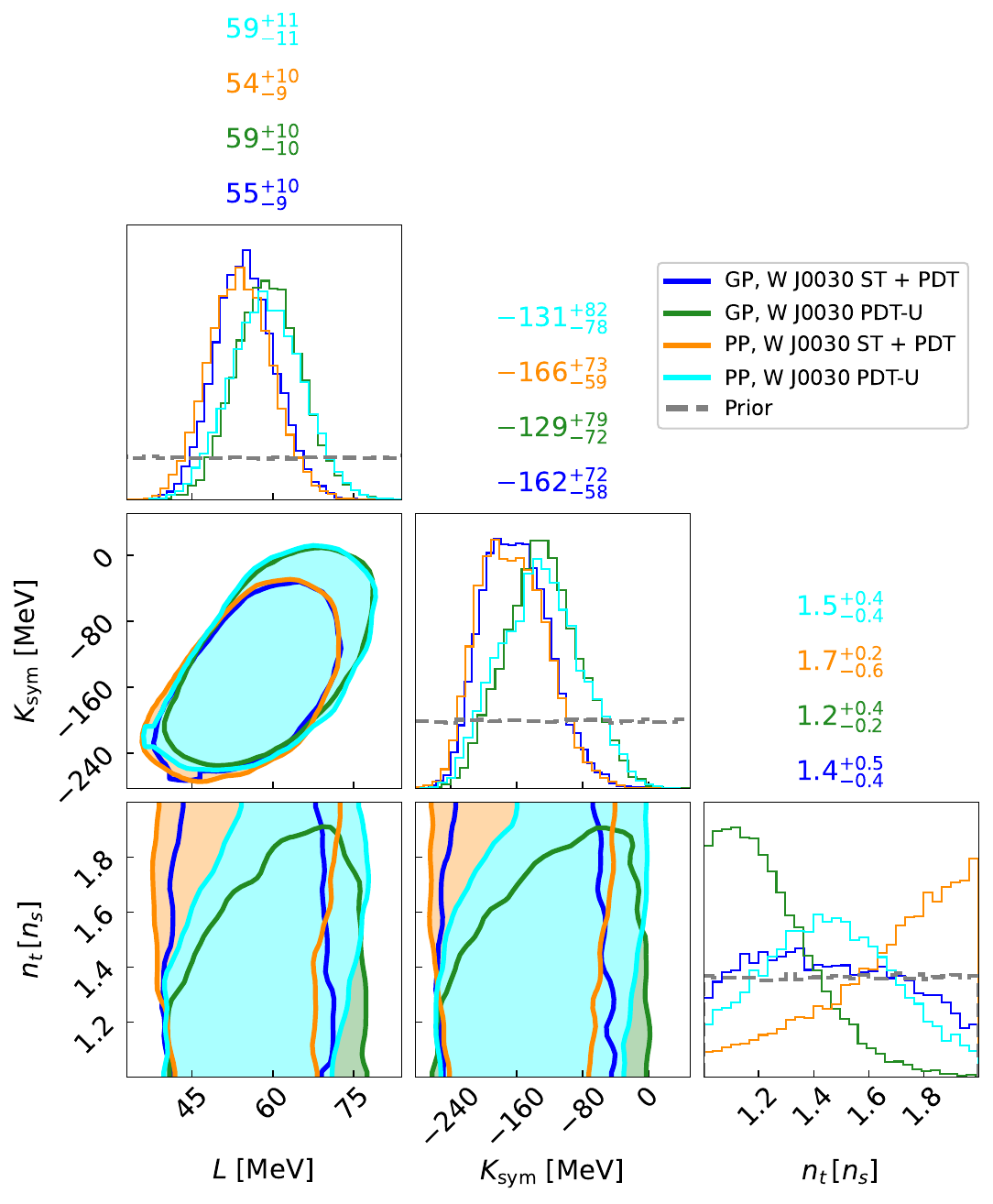}
\caption{Posterior distributions for the symmetry energy slope $L$, curvature $K_{\rm sym}$, and transition density $n_t$ under four scenarios: GP high-density extrapolation with J0030 ST+PDT (blue), GP with J0030 PDT-U (orange), PP with J0030 ST+PDT (green), and PP with J0030 PDT-U (red). The prior distribution is shown in grey.}
    \label{fig:common_eos_params}
\end{figure}

Figure~\ref{fig:common_eos_params} shows the posterior distributions of three key EOS parameters that are common across all scenarios: the symmetry energy slope $L$, the curvature of the symmetry energy $K_{\rm sym}$, and the transition density $n_t$ at which the empirical meta-model is matched to the high-density extension.  

A systematic trend emerges: across both EOS extrapolation strategies, adopting the PDT-U hotspot geometry systematically shifts the inferred values of $L$ and $K_{\rm sym}$ toward higher medians compared to the ST+PDT case. This effect is robust to whether the high-density EOS is modeled with a GP or PP extension, and the two extrapolation schemes provide broadly similar posteriors for $L$ and $K_{\rm sym}$.

The behavior of the transition density parameter $n_t$ highlights differences between the four scenarios. GP-based models tend to prefer somewhat lower transition densities compared to PP-based models, but the exact posteriors vary with the assumed hotspot geometry. This indicates that both the extrapolation strategy and the treatment of J0030’s pulse profile introduce distinct systematics into the inferred matching density.

\begin{figure}
    \centering
    \includegraphics[width=0.45\textwidth]{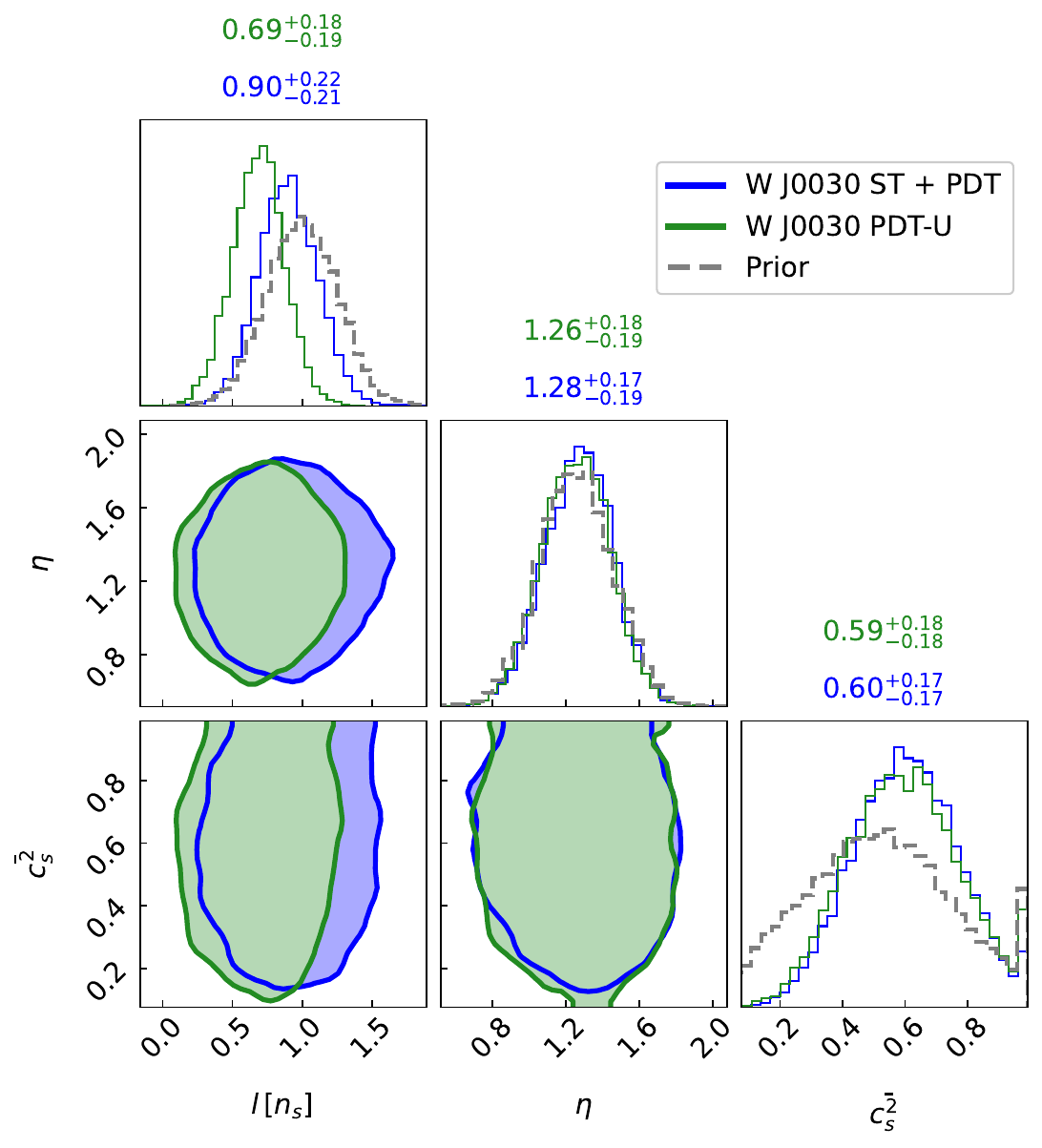}
    \caption{Posteriors of GP hyperparameters compared with priors. 
    $\eta$ remains prior dominated, $\bar c_s^2$ is robust across hotspot geometries, while $\ell$ shifts to smaller values in both cases, with PDT-U favoring the shortest correlation lengths.}

    \label{fig:GP_params}
\end{figure}

\begin{figure*}[t]
  \centering
  \includegraphics[width=\textwidth]{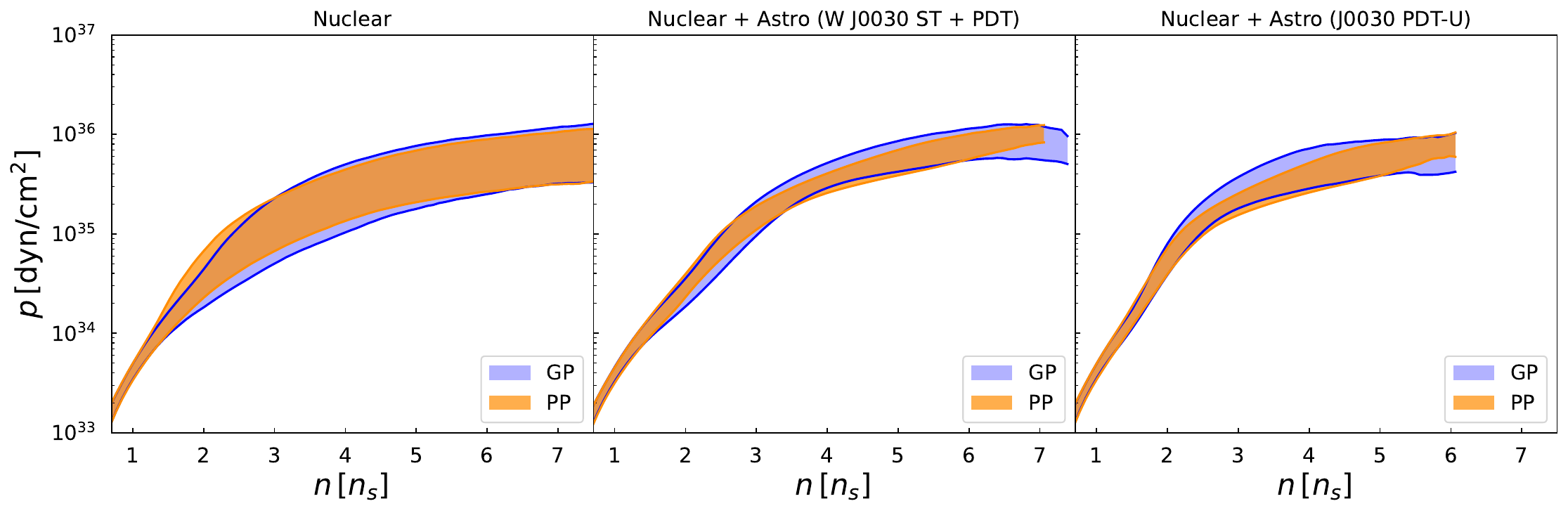}\\[6pt]
  \includegraphics[width=\textwidth]{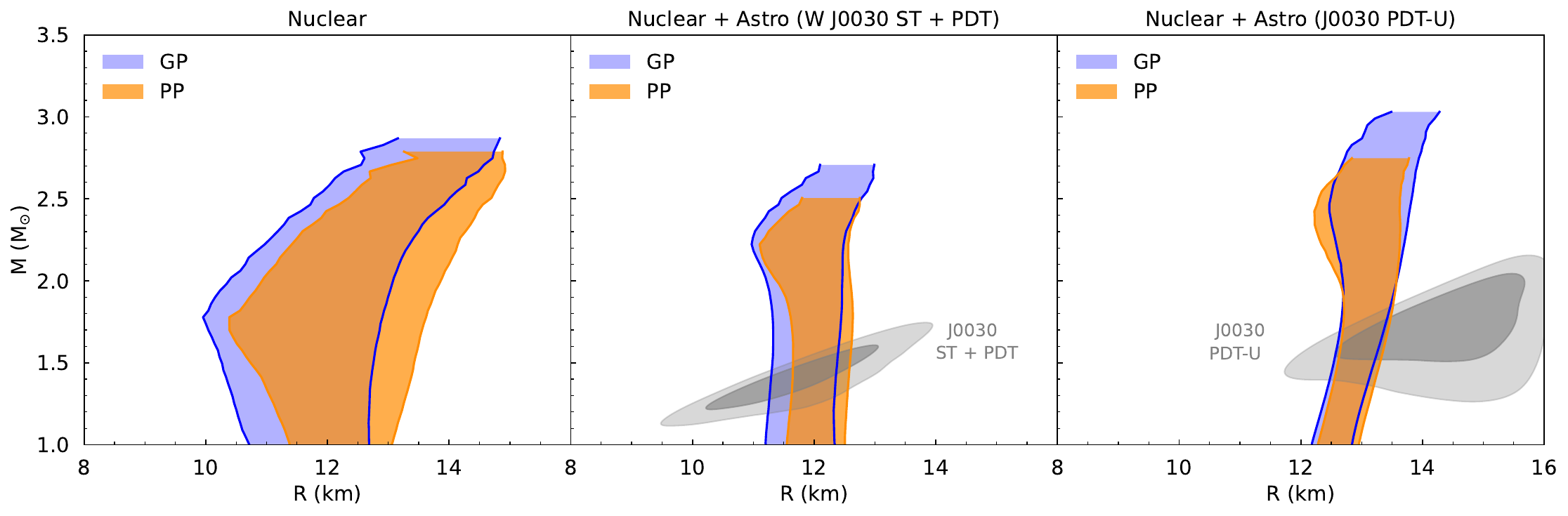}\\[6pt]
  \includegraphics[width=\textwidth]{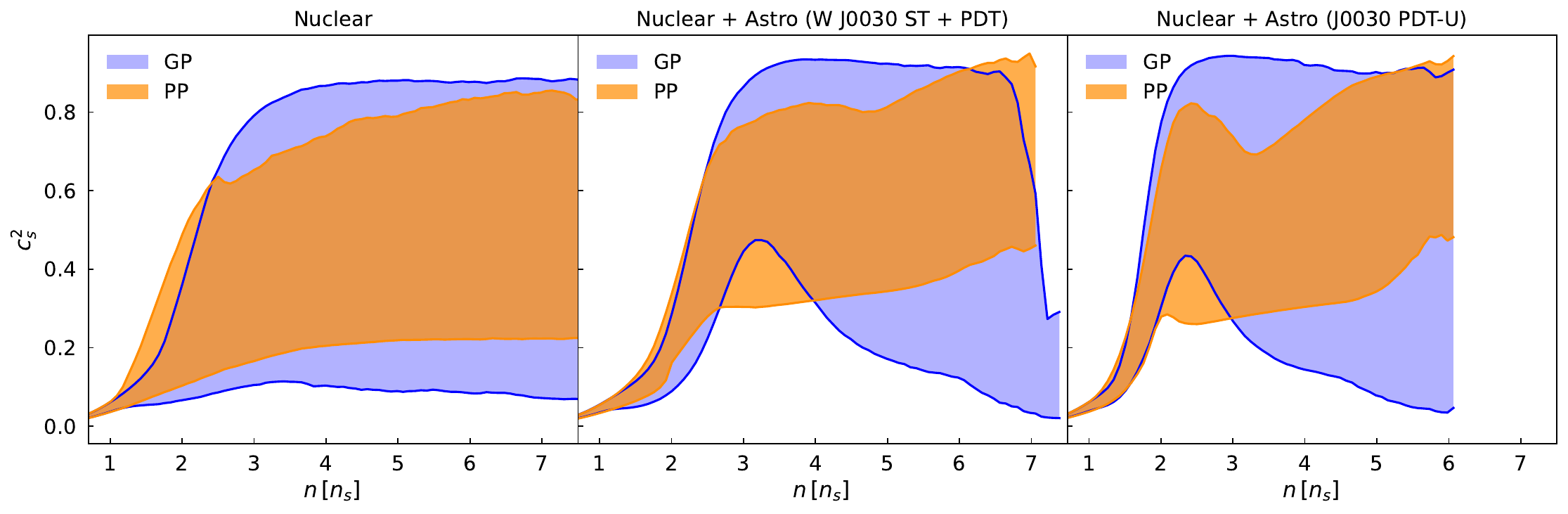}
  \caption{Posterior distributions of the EOS obtained with GP and 
    PP extrapolations under two hotspot geometries of 
    PSR~J0030+0451 (ST+PDT and PDT-U). Top: pressure--density relation. 
    Middle: mass--radius curves. Bottom: squared sound speed vs density.}
  \label{fig:three_plots}
\end{figure*}

\begin{figure*}[ht!]
    \centering
    \includegraphics[width=\textwidth]{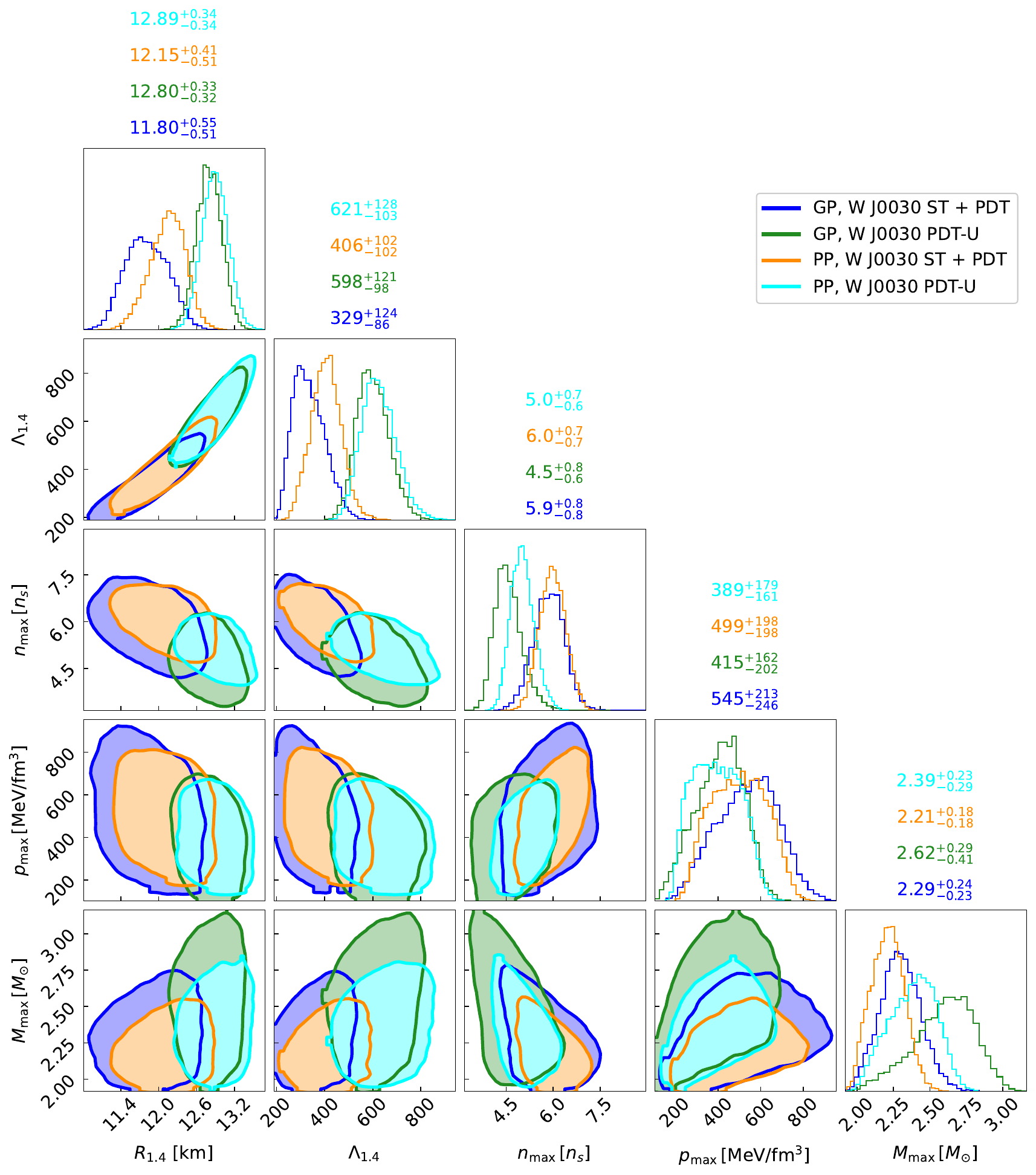}
\caption{Correlations between radius ($R_{1.4}$) and tidal deformability ($\Lambda_{1.4}$) of a $1.4M_{\odot}$ NS, maximum density ($n_{\rm max}$), maximum pressure ($p_{\rm max}$), and maximum mass ($M_{\rm max}$) are shown for each scenario considered in this study, as indicated in the legend of the plot. In the marginalized one-dimensional plots, the median and 90\% CIs are shown for each constraint type, with each constraint represented in its respective color.}
    \label{fig:macro_params}
\end{figure*}

\begin{table*}[htbp!]
    \centering
    \caption{Bayesian evidence comparison for EOS extrapolation models (PP vs.\ GP) and hotspot geometries (ST+PDT vs.\ PDT-U). 
    Quoted values are log-evidences with sampling uncertainties. 
    Evidence differences $\Delta \log Z$ are shown for both hotspot and extrapolation comparisons. 
    Interpretation follows Kass \& Raftery~\citep{bfactor}.}
    \label{tab:bayes_evidence}
    \begin{tabular}{lcccc}
        \hline\hline
        EOS Model & Geometry & $\log Z$ & $\Delta \log Z$ (Geom.) & $\Delta \log Z$ (EOS) \\
        \hline
        PP & ST+PDT & $-11.838   \pm 0.019$ & \multirow{2}{*}{$\approx +3.68$} (favor ST+PDT) & $\approx  -2.08$ (vs.\ GP) \\
        PP & PDT-U  & $-15.517   \pm 0.129$ &  & $\approx -1.24 $ (vs.\ GP) \\
        \hline
        GP & ST+PDT & $-9.754 \pm 0.015$   & \multirow{2}{*}{$\approx  + 4.52 $} (favor ST+PDT) & $\approx +2.08 $ (vs.\ PP) \\
        GP & PDT-U  & $-14.276 \pm 0.041 $ &  & $\approx  +1.24 $ (vs.\ PP) \\
        \hline\hline
    \end{tabular}
\end{table*}

\subsection{Constraints on GP hyperparameters}

Figure~\ref{fig:GP_params} shows the posterior distributions of the GP hyperparameters that regulate the high-density extrapolation of the EOS: the correlation length $\ell$, the variance amplitude $\eta$, and the reference mean squared sound speed $\bar c_s^2$. These are compared to the broad hierarchical priors described in Table~\ref{tab:prior}.  

The variance amplitude $\eta$ remains almost completely prior dominated. Its posterior closely follows the prior distribution, demonstrating that the current multimessenger dataset does not meaningfully constrain the typical size of local fluctuations in the squared sound speed.  

By contrast, the data strongly inform both $\ell$ and $\bar c_s^2$. The posteriors of $\bar c_s^2$ are consistently shifted relative to the prior, reflecting that the dataset places nontrivial constraints on the average stiffness of the EOS at supranuclear densities. Moreover, the two hotspot geometries for PSR~J0030+0451 yield nearly identical posteriors for $\bar c_s^2$, indicating that this parameter is robust against pulse-profile modeling choices.  

The correlation length $\ell$ exhibits more nuanced behavior. In both hotspot geometries, the posterior median of $\ell$ lies below the prior mean, suggesting that the data generally favor shorter correlation lengths than assumed a priori---that is, the EOS is allowed to vary less smoothly with density than the prior would indicate. Furthermore, a geometry dependence is evident: the PDT-U solution drives $\ell$ to even smaller values than the ST+PDT case. This preference can be interpreted as follows: since the PDT-U geometry infers larger neutron star radii, it requires a stiffer EOS at intermediate densities. Reconciling this with the remaining observational and theoretical inputs necessitates more localized adjustments in the sound-speed profile, which are enabled by reducing $\ell$. By contrast, the ST+PDT geometry is more compatible with the combined dataset and therefore favors smoother high-density extrapolations with moderately larger $\ell$.  

\subsection{Reconstructed high-density EOS and stellar structure under different modeling assumptions}

Figure~\ref{fig:three_plots} presents the inferred pressure--density relations (top), 
mass--radius curves (middle), and sound-speed profiles (bottom) for the four inference 
scenarios, comparing GP and PP extrapolations 
combined with either the ST+PDT or PDT-U hotspot geometry of PSR~J0030+0451.  

With only nuclear-theory and experimental likelihoods applied, the GP extrapolation 
already behaves differently from the PP case. GP EOSs favor systematically softer 
posteriors in the density range $1$--$3\,n_s$, while also spanning a visibly broader 
envelope across all densities. By contrast, the PP representation produces narrower 
credible regions, but at the cost of stronger dependence on its restrictive functional form.  

Adding astrophysical data (NICER, GW, and pulsar masses) dramatically improves the 
constraints. Under the ST+PDT geometry, the GP EOS posteriors are significantly softer 
than PP, yielding smaller radii, higher maximum masses, and larger central densities. 
The PDT-U geometry shifts the posteriors toward stiffer EOSs with correspondingly 
larger radii, but the differences between GP and PP extrapolations remain evident 
in the widths of their credible bands.  

The sound-speed reconstructions highlight these differences most clearly. For the 
ST+PDT case, the GP EOS produces somewhat tighter constraints on $c_s^2(n)$ up to 
$\sim 3.3\,n_s$, while for PDT-U the informative range extends only to 
$\sim 2.5\,n_s$. Moreover, the ST+PDT solution admits central densities higher 
than those of PDT-U, and the GP posteriors reveal a dip in $c_s^2$ near the 
central density in this case. Such features are inaccessible in the PP representation, 
which enforces stepwise continuity.  

\subsection{Macroscopic neutron star properties}

Figure~\ref{fig:macro_params} shows the posterior distributions of macroscopic neutron 
star properties inferred from our EOS models: the radius $R_{1.4}$ and tidal deformability 
$\Lambda_{1.4}$ of a $1.4\,M_\odot$ star, the maximum mass $M_{\max}$, the corresponding 
central density $n_{\max}$, and the central pressure $p_{\max}$. Results are compared across 
the four scenarios combining GP or PP extrapolations with the ST+PDT or PDT-U hotspot geometry 
of PSR~J0030+0451.  

For $R_{1.4}$ and $\Lambda_{1.4}$, both the extrapolation strategy and the hotspot geometry 
play important roles. Within the ST+PDT geometry, GP models predict systematically smaller 
radii and tidal deformabilities than PP models, with median shifts of about 
$\sim 350\,\mathrm{m}$ and $\sim 80$, respectively. Changing the hotspot geometry has a 
larger effect: moving from ST+PDT to PDT-U increases $R_{1.4}$ by roughly $\sim 1.1\,\mathrm{km}$ 
and $\Lambda_{1.4}$ by about $\sim 270$ in the GP case. While the precise values differ for PP, 
the same trend holds: PDT-U always favors larger radii and tidal deformabilities than ST+PDT.  

The maximum mass follows a consistent but distinct set of trends. PDT-U solutions yield 
larger $M_{\max}$ than ST+PDT in both GP and PP cases, reflecting their preference for stiffer 
EOSs. In addition, GP extrapolations systematically predict higher $M_{\max}$ than PP under 
the same geometry. This is in contrast to the behavior of $R_{1.4}$ and $\Lambda_{1.4}$, 
where GP tends to give smaller values than PP. The pattern highlights the flexibility of 
the GP representation: it can accommodate large maximum masses while allowing softer 
intermediate-density behavior.  

The corresponding central densities and pressures underscore these differences. 
ST+PDT models, which favor softer EOSs, are associated with higher $n_{\max}$ and $p_{\max}$ 
than PDT-U. This outcome is expected: softer EOSs must reach higher central densities and 
pressures to support massive stars, whereas stiffer EOSs achieve the same or larger 
$M_{\max}$ at lower central densities and pressures.  

Finally, we note that the posterior distributions of the neutron star mass distribution parameters are not shown here, as they are largely insensitive to the choice of hotspot geometry or EOS extrapolation scheme. These results are consistent with our previous analyses~\citep{Biswas:2024hja,Biswas:2025ivu}, where the corresponding distributions were presented in detail. For completeness, we simply refer the reader to those earlier works.

\subsection{Model Comparison via Bayesian Evidence}

We assess the relative support for different modeling assumptions using Bayesian evidence, 
considering both (i) the hotspot geometry in pulse-profile modeling (ST+PDT vs.\ PDT-U), 
and (ii) the high-density extrapolation scheme for the EOS (piecewise polytrope vs.\ Gaussian process). 
Following the interpretation of Kass and Raftery~\citep{bfactor}, 
we interpret differences in log-evidence in terms of the strength of model preference.  

For the PP model, the evidence difference between ST+PDT and PDT-U is 
\(\Delta \log Z \approx 3.68 \), corresponding to 
\(\log_{10}\mathrm{BF} \approx 1.59\). This constitutes \emph{strong evidence} in favor of ST+PDT.  
For the GP model, the preference is even stronger: 
\(\Delta \log Z \approx 4.52 \), or 
\(\log_{10}\mathrm{BF} \approx 1.96\), which constitutes \emph{very strong evidence}.  
Thus, under both extrapolation schemes, ST+PDT is favored, with the GP extension providing the sharper distinction.  

Comparing the extrapolation strategies themselves, the GP extension is preferred over PP in both hotspot geometries. 
For ST+PDT, the evidence difference is \(\Delta \log Z \approx  2.08 \), 
corresponding to a Bayes factor (BF) of about 8 (\(\log_{10}\mathrm{BF} \approx 0.9\)), 
indicating \emph{substantial evidence} for GP.  
For PDT-U, the difference is smaller, \(\Delta \log Z \approx 1.24 \), 
corresponding to a Bayes factor of about 3.5 (\(\log_{10}\mathrm{BF} \approx 0.54\)), 
indicating \emph{substantial evidence}.  

Overall, the dataset decisively favors the ST+PDT hotspot geometry, 
and within either geometry, shows a substantial preference for the GP extrapolation. 
The combination of ST+PDT+GP is the most strongly supported model.  

\subsection{Comparison with Ng et al.}  
\label{subsec:ng-comparison}

Our results can be compared directly with the recent work of Ng et al.~\citep{Ng:2025wdj}, who also adopt a semi-nonparametric EOS framework based on Gaussian Processes. Using a comparable set of astrophysical observations, they report $R_{1.4} = 11.4^{+0.98}_{-0.60}$ km and $M_{\max} = 2.31^{+0.35}_{-0.23} M_\odot$ (90\% C.L.), whereas our ST+PDT+GP analysis yields $R_{1.4} = 11.8^{+0.55}_{-0.51}$ km and $M_{\max} = 2.29^{+0.24}_{-0.23} M_\odot$. Although the median values are consistent, the credible intervals reported by Ng et al. are noticeably broader than those obtained in this work.  

An important difference lies in the inference methodology. Ng et al. do not provide explicit details of their sampling scheme. However, their Gaussian-process framework closely follows Landry et al.~\citep{Landry:2018prl} and Essick et al.~\citep{Essick:2019ldf}, where posterior distributions are obtained via direct Monte Carlo draws of EOS realizations from the GP prior, reweighted by the likelihood. Convergence in that approach is typically monitored using the effective sample size. While this procedure is formally correct, it is computationally inefficient in high-dimensional problems where the posterior occupies only a small fraction of the prior volume. As a result, many samples carry negligible weight, limiting the effective number of posterior draws and leading to broader credible intervals.  

By contrast, our analysis employs nested sampling as implemented in {\tt PyMultiNest}, which adaptively explores the high-likelihood regions of parameter space and simultaneously yields Bayesian evidence estimates. This strategy achieves a more efficient exploration of the posterior distribution, resulting in tighter constraints on $R_{1.4}$ and $M_{\max}$, while also enabling robust model comparison between different EOS extrapolation strategies and hotspot geometries. We therefore attribute part of the broader uncertainties in Ng et al. to the sampling methodology, in addition to possible differences in prior choices and observational inputs.

\section{Conclusion}  
\label{sec:conclusion}  

In this work, we have extended our earlier hybrid EOS framework~\cite{Biswas:2024hja} by replacing the high-density piecewise-polytropic extrapolation with a Gaussian Process representation of the squared sound speed. This allows a systematic comparison between parametric and nonparametric extrapolation strategies while retaining the same physics-informed SLy crust and nuclear meta-model near saturation density.  

Using a hierarchical Bayesian framework that incorporates constraints from NICER, gravitational waves, radio pulsar mass measurements, neutron-skin thickness experiments, and theoretical input from $\chi$EFT and pQCD, we performed joint inference of the neutron star EOS and mass distribution under four scenarios: two EOS extrapolations (PP and GP) and two hotspot geometries for PSR J0030+0451 (ST+PDT and PDT-U).  

Our previous works~\cite{Biswas:2024hja,Biswas:2025ivu} demonstrated that hotspot geometry assumptions strongly influence EOS inferences when using piecewise-polytropic extrapolations, with the PDT-U model favoring larger radii and stiffer EOSs than the ST+PDT configuration. Here we show that the same qualitative behavior persists with the GP extension, confirming that this effect is robust across extrapolation strategies. Quantitatively, GP-based EOSs yield broader posteriors than PP models, but generally predict higher maximum masses and smaller neutron star radii. Bayesian model comparison strongly favors the ST+PDT geometry over PDT-U in both extrapolation schemes, and---while not decisive overall---provides substantial support for the GP extension over the PP model, particularly in the ST+PDT case.  

Overall, our results highlight that both astrophysical systematics (hotspot geometry assumptions) and theoretical systematics (EOS extrapolation schemes) continue to play a central role in dense-matter inference. Looking forward, the evidence in favor of the GP framework suggests that nonparametric extrapolations may offer a more reliable path toward robust, bias-minimized EOS constraints in the multimessenger era.

\section*{Acknowledgements}
  BB thanks the anonymous referees for useful suggestions which helped to improve the manuscript. BB  acknowledges the support from the Alexander von Humboldt Foundation through a Humboldt Research Fellowship for Postdoctoral Researchers. Calculations were performed on the facilities at the SUNRISE HPC facility supported by the Technical Division at the Department of Physics, Stockholm University. 

\bibliography{mybiblio}
\end{document}